\documentstyle[a4,12pt]{article}
\newcommand{\eps}{\epsilon}
\newcommand{\pdt}{\partial_t}
\newcommand{\pdx}{\partial_x}
\newcommand{\pdxd}{\partial_{x'}}
\newcommand{\pdy}{\partial_y}

\newcommand{\dxi}{\partial_\xi}
\newcommand{\deta}{\partial_\eta}

\begin{document}
\title{Renormalization Group Method and Reductive Perturbation Method}
\author{Ken-ichi Matsuba and Kazuhiro Nozaki\\
Department of Physics,Nagoya University,Nagoya 464-01,Japan}
\maketitle
\begin{abstract}
It is shown that the renormalization group method 
does not necessarily eliminate all secular terms in perturbation series 
 to partial differential equations
and a functional subspace of renormalizable secular solutions
 corresponds to a choice of scales of independent variables in 
 the reductive perturbation method.
\end{abstract}
Recently a novel method based on the perturbative renormalization group
theory has been developed as an asymptotic singular 
perturbation technique by L.Y.Chen, N.Goldenfeld and Y.Oono \cite{oono1}
and the usefulness of the method has been amply demonstrated \cite{oono2}.
Their renormalization group method (the RG method) removes 
secular or divergent terms from a perturbation series 
by renormalizing integral constants
of  lower order solutions. 
It is a crucial procedure to
obtain secular solutions explicitly by perturbative analysis,
which is usually easy and clear for 
ordinary differential equations (ODE) and the method
has made impressive success in application to ODE \cite{oono2}.
However, applying the RG method to partial differetial equations (PDE),
it should be noted that all of secular solutions to PDE  
can not be obtained unless a functional space
of secular solutions is specified.  
This point has not been discussed 
explicitly in previous application of the RG method to PDE.\par  
It may be natural to restrict a functional space of secular solutions
to a family of polynomial-type functions
of independent variables. 
 We shall show some physical examples where 
 all of secular solutions of polynomial-type
  can not be removed by the renormalization procedure and
  we must impose further restrictions on the functional space 
  in order to remove secular solutions. 
  Unless a functional space
of secular solutions is specified,
  the RG method does not necessarily yield the unique 
  renormalization group (RG) equation
 in  application to PDE . \par
The purpose of this letter is to show through some examples that            
 a functional subspace of secular solutions
 corresponds to a choice of scales of independent variables in 
 the reductive perturbation (RP) method \cite{tani}.\par
(1) As the first example, let us derive the nonlinear Schr\"odinger 
equation from the following simple wave equation:
\begin{equation}
\pdt^2u-\pdx^2u+(1+au^2)u=0,\label{nwe}
\end{equation}
where $a$ is constant. Substituting the expansion
\begin{equation}
u=\eps u_1+\eps^2 u_2+\eps^3 u_3+\cdots,\label{exp}
\end{equation}
where $\eps$ is a small parameter, into Eq.(\ref{nwe}), we have
\begin{eqnarray}
u_1&=&A\exp[i(kx-\omega t)]+c.c.,\label{u1}\\
u_2&=&0,\label{u2}
\end{eqnarray}
\begin{equation}
(\pdt^2-\pdx^2+1)u_3=-3a|A|^2A \exp[i(kx-\omega t)],\label{lwe}
\end{equation}
where $\omega=\sqrt{1+k^2}$, $c.c.$ denotes complex conjugate
 and only a singular term is
 retained in the right-hand side of Eq. (\ref{lwe}).  
It may be plausible to set a secular solution of Eq. (\ref{lwe})
 in the following polynomial type:
$$u_3=P(x,t)(-3a)|A|^2A \exp[i(kx-\omega t)],$$
where $P$ is a polynomial of $x$ and $t$.
Then, Eq. (\ref{lwe}) is rewritten as
\begin{eqnarray}
LP&=&1,\label{p1}\\
L&\equiv&\pdt^2-2i\omega\pdt-\pdx^2-2ik\pdx.\nonumber
\end{eqnarray}
We obtain four polynomial solutions of Eq. (\ref{p1}):
\begin{eqnarray*}
P_1&=&it/2\omega,\quad P_2=ix/2k,\\
P_3&=&-\omega^2x '^2/2,\\
P_4&=&-(\omega/2k)(x ' t-i\omega^3x '^3/3),
\end{eqnarray*}
where $x '\equiv x-kt/\omega$. The other solutions are expressed by
 adding polynomials belonging to a kernel of the operator $L$ 
 but not being linear combinations of $P_j (j=1,2,3,4)$ to 
 the four solutions. Since secular terms belonging to
  a kernel of $L$ have not an effect on the RG equation,
we do not pay attension to such difference in secular solutions.
A secular solution $P_4$  can not apparently be removed
by the usual renormalization procedure (not renormalizable)
 and we obtain a functional space of renormalizable 
 secular solutions, which consists of
$P_j (j=1,2,3)$. Then, renormalizable secular solutions are given as
$P=c_1P_1+c_2P_2+c_3P_3, c_1+c_2+c_3=1$ or, in terms of $(x ',t)$,
\begin{equation} 
P=(c_1+c_2)\frac{it}{2\omega}-c_3\frac{\omega^2}{2}x '^2
+c_2\frac{ix '}{2k},
\label{scsl1}
\end{equation}
where the last term in Eq. (\ref{scsl1}) belongs to a kernel
of $L$ and does not have an effect on the RG equation. 
The secular solutions (\ref{scsl1}) are removed by renormalizing 
the first-order complex amplitude $A$ through the renormalization 
procedure introducd in \cite{oono1}.  Thus, we obtain 
the nonlinear Sch\"odinger (NS) equation as a RG equation:
$$i\pdt A+\frac{1}{2\omega^3}\pdxd^2A-
\eps^2\frac{3a}{2\omega}|A|^2A=0,$$ 
 which gives  well-known scaling $x'\to \eps x'$ and $t\to \eps^2 t$
 of the NS equation  in the RP method \cite{tani}.
It should be noted that the NS equation is not the unique
 RG equation in the present case.
 If we choose a subspace consisting of $P_1$ and $P_2$
as a functional space of renormalizable secular solutions,
we have another RG equation:
$$i(\pdt +\frac{k}{\omega}\pdx)A-
\eps^2\frac{3a}{2\omega}|A|^2A=0,$$ 
which corresponds to scaling $x\to \eps^2x$ and $t\to \eps^2 t$ 
in the RP method.\par   
(2) Let us derive a slow amplitude equation from
the following model equation:
\begin{equation}
\{\pdt[\pdt+(k^2+\triangle)^2-\eps^2+u^2]-\triangle\}u
+(1+au^2)u=0,\label{hpb}
\end{equation}
where $\eps$ is a small bifurcation parameter and 
$\triangle\equiv\pdx^2+\pdy^2$.
The nonlinear wave equation (\ref{nwe}) and 
the Swift-Hohenberg equation \cite{swft} are combined in 
the model equation (\ref{hpb}) ,which 
may be the simplest equation describing the Hopf
bifurcation in continuous media. 
Expanding $u$ in the same form as Eq. (\ref{exp}),
we have $u_1$ and $u_2$  given in Eqs. (\ref{u1}) and (\ref{u2}) 
respectively. The third-order secular correction obeys
\begin{equation}
\{\pdt[\pdt+(k^2+\triangle)^2]-\triangle+1\}u_3
=[-i\omega+3(i\omega-a)|A|^2]A\exp[i(kx-\omega t)].\label{lhpb}
\end{equation}
Let us seek secular solutions of Eq.(\ref{lhpb}) in the form
$$u_3=P(x,y,t)[-i\omega+3(i\omega-a)|A|^2]A\exp[i(kx-\omega t)],$$
then we have
\begin{eqnarray}
LP&=&1,\label{p2}\\
L&\equiv&\pdt[\pdt-2i\omega+(\triangle+2ik\pdx)^2]
-i\omega(\triangle+2ik\pdx)^2-(\triangle+2ik\pdx).\nonumber
\end{eqnarray}
For simplicity, we list only six solutions of lower power 
among sixteen polynomial solutions of Eq.(\ref{p2}).
\begin{eqnarray*}
P_1&=&\frac{it}{2}\omega,\quad P_2=\frac{ix}{2k},\\
P_3&=&\frac{-\omega^2}{2-8i\omega^3k^2}x'^2,\\
P_4&=&-\frac{y^2}{2},\\
P_5 &=& -\frac{\omega(1-4iw^3k^2)}{2k(1+4iw^3 k^2+4iw^3)}(x't-\frac{iw^3}{3(1-4iw^3k^2)}x'^3),\\
P_6 &=& \frac{w}{48k^3(1+2iw^3)}(x'^3-3\frac{1-4iw^3k^2}{w^2}x'y^2).\\
\end{eqnarray*}
The other solutions of higher power are not renormalizable and 
renormalizable secular solutions are given, in terms of $(x ',t)$,
by 
$$ 
P=(c_1+c_2)\frac{it}{2\omega}-c_3\frac{\omega^2}{2-8i\omega^3k^2}x '^2
-c_4\frac{y^2}{2}+c_2\frac{ix '}{2k},
$$
where $\sum\nolimits_{j=1}^4c_j=1$.
Then, the RG equation becomes
\begin{equation}
i\pdt A+\frac{1-4i\omega^3k^2}{2\omega^3}\pdxd^2A
+\frac{1}{2\omega}\pdy^2A
+\eps^2[-\frac{i}{2}+3(\frac{i}{2}-\frac{a}{2\omega})|A|^2]A
=0,\label{cgl}
\end{equation}
which is the two-dimensional complex Ginzburg-Landau equation.
>From Eq. (\ref{cgl}), we have scaling $x'\to\eps x',y\to\eps y$ and
$t\to\eps^2 t$ in the RP method \cite{matsu}.
When we choose another subspace of renormalizable secular solutions
such as $c_1P_1+c_2P_2+c_4P_4$, we obtain
$$i(\pdt+\frac{k}{\omega}\pdx)A
+\frac{1}{2\omega}\pdy^2A
+\eps^2[-\frac{i}{2}+3(\frac{i}{2}-\frac{a}{2\omega})|A|^2]A
=0,$$
which corresponds to scaling $x\to\eps^2 x, y\to\eps y$ and
$t\to\eps^2 t$ in the RP method. 
It should be noted that the Newell-Whitehead (NW) scaling 
$x\to\eps x, y\to\eps^\frac{1}{2} y$ and $t\to\eps^2 t$ 
 can not be reduced
 whichever subspace of renormalizable secular solutions 
 is chosen. This fact indicates the reason why the NW scaling 
 yields such an inconsistent equation as a generalized 
 Newell-Whitehead-Segel equation introduced in the
 case of the Hopf bifurcation  \cite{brand}.   
 \par
 (3) As a final example, we derive 
 the Kadomtsev-Petviashvili (KP) equation from 
 the following weakly dispersive nonlinear wave equation:
 \begin{equation}
 \pdt^2 u-\nabla\cdot[(1+au)\nabla u]+\eps b\triangle^2 u=0,\label{nw2}
 \end{equation}
 where $\nabla$ is the two-dimensional gradient operator; 
 $a$ and $b$ are constant. Substitutig $u=\eps u_1+\eps^2u_2+\cdots$ in
 Eq. (\ref{nw2}), we get, to order $\eps^2$,
 \begin{eqnarray}
 (\pdt^2-\triangle)u_1&=&0,\label{lnw1}\\
 (\pdt^2-\triangle)u_2&=&a\nabla\cdot(u_1\nabla u_1)-b\triangle^2 u_1.
 \label{lnw2}
 \end{eqnarray}   
 Let us choose a plane-wave solution of Eq. (\ref{lnw1}) as
 $u_1=f(\xi), \xi=x-t,\quad$  then Eq. (\ref{lnw2}) reads
 \begin{equation}
 (4\dxi\deta+\pdy^2)u_2=-\dxi F(\xi),\label{lnw21}
 \end{equation}
 where $\eta=x+t,\quad F=af\dxi f-b\dxi^3 f$. Possible secular solutions 
 to Eq. (\ref{lnw21}), except a kernel of the operator $4\dxi\deta+\pdy^2$, 
   are given by
\begin{eqnarray}
u_2&=&-\frac{c_1}{4}F\eta-\frac{c_2}{2}\dxi Fy^2,\nonumber\\
   &=&-\frac{1}{2}(c_1Ft+c_2\dxi Fy^2)-\frac{c_1}{4}F\xi,\label{scs2}
\end{eqnarray}
where $c_1+c_2=1$. The last term in Eq. (\ref{scs2}) is non-secular
when  $F\xi$ is assumed to be finite as $|\xi| \to\infty$.
Eliminating secular terms in Eq. (\ref{scs2}), we obtain 
the KP equation
$$
(\pdt\dxi+\frac{1}{2}\pdy^2)f=-\frac{\eps}{2}\dxi(af\dxi f-b\dxi^3 f),
$$
from which  KP scaling $t\to\eps^{\frac{3}{2}}t, \xi\to\eps^{\frac{1}{2}}
\xi,y\to\eps y$ is reduced. If we set $c_2=0$ in Eq. (\ref{scs2}),
we get the K-dV equation as a RG equation.
Since Eq. (\ref{nw2}) is simple, 
all possible secular solutions
 to order $\eps^2$ happen to be eliminated in this example.
 \par
 In summary,  
the RG method does not necessarily eliminate 
all secular terms in perturbation series to PDE
and relates to the RP method in the manner that
 a functional subspace of renormalizable secular solutions in
 the RG method corresponds to a choice of scales of 
 independent variables in the RP method. Therefore, 
 the RG method does not necessarily yield the unique RG equation
 in application to PDE.

\end{document}